\documentclass[aps,twocolumn,showpacs,groupedaddress]{revtex4-1}
\usepackage{amssymb}
\usepackage{mathrsfs}
\usepackage{pstricks}
\usepackage{color}
\usepackage{graphicx}
\usepackage[fleqn]{amsmath}
\usepackage{slashed}
\usepackage{amsmath}
\usepackage{mathtools}
\usepackage{array}
\usepackage{tabularx}
\usepackage{physics}
\usepackage{slashed}
\usepackage[utf8]{inputenc}
\usepackage[T1]{fontenc}
\usepackage{float}
\begin{document}

%\preprint{}

\title{\bf Is $\gamma_{KLS}$-generalized statistical field theory complete?}


%\author{William C. Vieira}
%\email{william.vieira@gmail.com}
%\affiliation{\it Departamento de F\'\i sica, Universidade Federal do Piau\'\i, 64049-550, Teresina, PI, Brazil}
%\author{T. F. A. Alves}
%\author{J. F. da Silva Neto}
%\author{F. W. S. Lima}
\author{Paulo R. S. Carvalho}
\email{prscarvalho@ufpi.edu.br}
\affiliation{\it Departamento de F\'\i sica, Universidade Federal do Piau\'\i, 64049-550, Teresina, PI, Brazil}
%\author{G. A. Alves}
%\affiliation{\it Departamento de F\'{i}sica, Universidade Estadual do Piau\'{i}, 64002-150, Teresina - PI, Brazil}

%\author{T. F. A. Alves}
%\author{Paulo R. S. Carvalho}
%\email{prscarvalho@ufpi.edu.br}
%%\address[1]{\it Departamento de F\'{i}sica, Universidade Federal do Piau\'{i}, 64049-550, Teresina, PI, Brazil}
%\affiliation{\it Departamento de F\'\i sica, Universidade Federal do Piau\'\i, 64049-550, Teresina, PI, Brazil}

%\thanks{}


%\date{\today}

\begin{abstract}
In this Letter we introduce some field-theoretic approach for computing the critical properties of $\gamma_{KLS}$-generalized systems undergoing continuous phase transitions, namely $\gamma_{KLS}$-statistical field theory. From this new approach emerges the new generalized O($N$)$_{\gamma_{KLS}}$ universality class, which is capable of encompassing nonconventional critical exponents for real imperfect systems known as manganites not described by standard statistical field theory. We compare the generalized results with those obtained from measurements in manganites. The agreement was satisfactory, where the relative errors are $< 5\%$ for the most of manganites used. Although the present approach describes the aforementioned nonconventional critical indices, we show that it is not complete. For example, it does not explain the results for some other manganites, being explained only for nonextensive statistical field theory recently introduced in literature. So, $\gamma_{KLS}$-statistical field theory has to be discarded for statistical mechanics generalization purposes.
\end{abstract}

%\pacs{11.30.-j; 64.60.ae; 11.30.Cp}
%\keywords{}

\maketitle


\section{Introduction}

\par Recently, a field-theoretic renormalization group method for describing the unconventional critical behavior of some real imperfect systems known as manganites \cite{Magnetochemistry.Turki,KHELIFI2014149,PhysRevB.75.024419,OMRI20123122,Ghosh_2005,doi:10.1063/1.2795796,GHODHBANE2013558,J.Appl.Phys.A.Berger,PhysRevB.68.144408,BenJemaa,PhysRevB.70.104417,PhysRevB.79.214426,J.Appl.Phys.Vasiliu-Doloca,YU2018393,PhysRevB.92.024409,ZHANG2013146,PHAN201440,RSCAdvJeddi,HCINI20152042,Phys.SolidStateBaazaoui} was proposed \cite{CARVALHO2023137683}. These systems are complex and present a large number of strong interacting particles, defects, impurities, inhomogeneities, nonlinearities, competition etc. Such systems can not be understood by employing standard Gibbs-Boltzmann statistical field theory (SFT) formulated by Kenneth Wilson \cite{PhysRevLett.28.240}. Rather, we have to employ some generalized version of Wilson field-theoretic renormalization group, namely nonextensive statistical field theory (NSFT) \cite{CARVALHO2023137683}. As it is known, there are many attempts of generalizing Gibbs-Boltzmann statistical mechanics \cite{Tsallis1988,KANIADAKIS2001405,BECK2003267,PhysRevE.71.046128,Eur.Phys.J.B70.3}. In this direction, we have to check what of these proposed generalizations satisfy to the all requirements needed for a consistent generalized theory. Some of these conditions are: they have to emerge from a maximum principle and a trace-form entropy, present decisivity, maximality, concavity, Lesche stability, positivity, continuity, symmetry and expansibility. Once all these conditions are satisfied, the corresponding proposed generalized theory has to be applied to all experimental situations not described by the standard theory. If there is, at least, one experimental situation in which some proposed generalized statistics does not describe the referred experimental results, this statistics is not general enough to represent a generalized theory. Then it has to be discarded as one candidate to generalize statistical mechanics. In fact, recently it was shown that it was the case for Kaniadakis attempt for generalizing statistical mechanics \cite{ALVES2023138005}, \emph{i. e.}, this statistics, namely Kaniadakis statistics \cite{KANIADAKIS2001405} does not describe the set of unconventional critical exponents values for some manganites as NSFT does \cite{CARVALHO2023137683}. 

\par The aim of this Letter is to search if there are some field-theoretic renormalization group generalizations satisfying all the requirements aforementioned other than NSFT (parameterized by the parameter $q$ encoding some effective interaction \cite{CARVALHO2023137683}). For that, we employ the $\gamma_{KLS}$-generalized statistics proposed in Ref. \cite{PhysRevE.71.046128}. It is parameterized by the parameter $\gamma$ \cite{PhysRevE.71.046128,Eur.Phys.J.B70.3}. In the present context of critical exponents, we have to avoid to represent different concepts with the same letter, for example the $\gamma$ parameter and the susceptibility critical exponent $\gamma$. Then we have to use $\gamma_{KLS}$ instead $\gamma$ for representing that parameter. We will call such a theory as $\gamma_{KLS}$-generalized statistical field theory or $\gamma_{KLS}$-SFT for short. The range of $\gamma_{KLS}$ is $-1/2 < \gamma_{KLS} < 1/2$ \cite{PhysRevE.71.046128,Eur.Phys.J.B70.3}. We expect to recover the nongeneralized critical exponents values obtained by Kenneth Wilson \cite{PhysRevLett.28.240} in the limit $\gamma_{KLS}\rightarrow 0$. Also we expect the emergence of some $\gamma_{KLS}$-generalized universality class for $\gamma_{KLS}$-generalized Ising-like systems, namely the O($N$)$_{\gamma_{KLS}}$ one. Now the critical indices will depend on the dimension $d$, $N$ and symmetry of some $N$-component order parameter, if the interactions of the constituents of the systems are of short- or long-range type and $\gamma_{KLS}$. From the physical interpretation of the results, it will emerge the corresponding physical interpretation of the $\gamma_{KLS}$ parameter. For a similar field-theoretic approach, see Ref. \cite{1674-1137-42-5-053102}.

\section{$\gamma_{KLS}$-SFT}

\par We introduce the $\gamma_{KLS}$-SFT by defining its generating functional by
\begin{eqnarray}\label{huyhtrjisd}
&& Z[J] = \mathcal{N}^{-1}\exp_{\gamma_{KLS}}\left[-\int d^{d}x\mathcal{L}_{int}\left(\frac{\delta}{\delta J(x)}\right)\right]\times \nonumber \\&& \int\exp\left[\frac{1}{2}\int d^{d}xd^{d}x^{\prime}J(x)G_{0}(x-x^{\prime})J(x^{\prime})\right]
\end{eqnarray}
where 
\begin{eqnarray}
&& e_{\gamma_{KLS}}^{-x} =  \left[\left(\frac{1 + \sqrt{1 + 4\gamma_{KLS}^{3}x^{3}}}{2}\right)^{\frac{1}{3}} + \right.  \nonumber \\&&  \left. \left(\frac{1 - \sqrt{1 + 4\gamma_{KLS}^{3}x^{3}}}{2}\right)^{\frac{1}{3}}\right]^{\frac{1}{\gamma_{KLS}}},
\end{eqnarray}
is the $\gamma_{KLS}$-exponential function \cite{PhysRevE.71.046128,Eur.Phys.J.B70.3}. We can determine the constant $\mathcal{N}$ from the condition $Z[J=0] = 1$. Now we can compute the static $\gamma_{KLS}$-generalized critical exponents for O($N$)$_{\gamma_{KLS}}$ universality class for $\gamma_{KLS}$-$\phi^{4}$ theory through six distinct and independent methods in $d = 4 - \epsilon$ dimensions. The two independent $\gamma_{KLS}$-generalized indices valid for all loop levels are given by
\begin{eqnarray}\label{etaphi4}
\eta_{\gamma_{KLS}} = \eta + \frac{\gamma_{KLS}}{1 - \gamma_{KLS}}\frac{(N + 2)\epsilon^{2}}{2(N + 8)^{2}},  
\end{eqnarray}
\begin{eqnarray}\label{nuphi4}
\nu_{\gamma_{KLS}} = \nu + \frac{\gamma_{KLS}}{1 - \gamma_{KLS}}\frac{(N + 2)\epsilon}{4(N + 8)}.
\end{eqnarray}
The $\eta$ and $\nu$ indices are the corresponding nongeneralized critical indices valid for all loop orders. The corresponding dynamic $\gamma_{KLS}$-generalized critical index is given by 
\begin{eqnarray}\label{z}
z_{\gamma_{KLS}} = z + \frac{\gamma_{KLS}}{1 - \gamma_{KLS}}\frac{[6\ln(4/3) - 1](N + 2)}{2(N + 8)^{2}}  \epsilon^{2},
\end{eqnarray}
where $z$ is the nongeneralized index value valid for all loop levels.

\section{Comparison between theoretic and experimental results}

\par We compare the results of Eqs. (\ref{etaphi4})-(\ref{nuphi4}) and scaling relations among them with those obtained from experiments for some manganites in Tables \ref{tableexponentsE3N3}-\ref{tableexponentsE3N1}. We use the values $\eta = 0.061(8)$, $\nu = 0.689(2)$, $\beta = 0.365(3)$, $\gamma = 1.336(4)$ \cite{PHAN2010238} and $\eta = 0.030(4)$, $\nu = 0.630(1)$, $\beta = 0.325(2)$, $\gamma = 1.241(2)$ \cite{PHAN2010238} for nongeneralized Heinsenberg and Ising, respectively. \par We observe in Tables \ref{tableexponentsE3N3}-\ref{tableexponentsE3N1} that experimental values obtained from experiments for various manganites ($\gamma_{KLS}\neq 0$) are distinct from those nongeneralized ones ($\gamma_{KLS} = 0$). The agreement is satisfactory, withing some margin of error of $< 5\%$ for the most of manganites. Then, the critical behavior of manganites is nonconventional \cite{Magnetochemistry.Turki,KHELIFI2014149,PhysRevB.75.024419,OMRI20123122,Ghosh_2005,doi:10.1063/1.2795796,GHODHBANE2013558,J.Appl.Phys.A.Berger,PhysRevB.68.144408,BenJemaa,PhysRevB.70.104417,PhysRevB.79.214426,J.Appl.Phys.Vasiliu-Doloca,YU2018393,PhysRevB.92.024409,ZHANG2013146,PHAN201440,RSCAdvJeddi,HCINI20152042,Phys.SolidStateBaazaoui} and can not be described by conventional SFT \cite{PhysRevLett.28.240}. They belong to both $\gamma_{KLS}$-Heinsenberg ($N = 3$) and  $\gamma_{KLS}$-Ising ($N = 1$) universality classes, respectively. The $\gamma_{KLS}$-generalizaed critical exponents (temperature) values increase (decrease, see Table \ref{tablecriticalTC}) with increasing of $\gamma_{KLS}$.  

\begin{table}[H]
\caption{Static $\gamma_{KLS}$-generalized critical exponents ($\gamma_{KLS} \neq 0$) to $3$d ($\epsilon = 1$) Heisenberg ($N = 3$) systems, obtained from experiment through Modified Arrott (MA) plots \cite{PhysRevLett.19.786}, Kouvel-Fisher (KF) method \cite{PhysRev.136.A1626} and $\gamma_{KLS}$-SFT of this work.}
\begin{tabular}{ p{5.7cm}p{1.3cm}p{1.2cm}  }
 \hline
 %\multicolumn{4}{|c|}{Country List} \\
 \hline
 $\gamma_{KLS}$-Heisenberg & $\beta_{\gamma_{KLS}}$ & $\gamma_{\gamma_{KLS}}$    \\
 \hline
 %\multicolumn{4}{|c|}{Country List} \\
% \hline
% La$_{0.8}$Ca$_{0.2}$Mn$_{0.7}$Co$_{0.3}$O$_{3}$\cite{Magnetochemistry.Turki}MAP  &   0.333(30) &  1.298(20)  \\
% $q$ = 1.5 This work  &  0.343(3) &  1.267(7)     \\
% \hline
 La$_{0.67}$Sr$_{0.33}$MnO$_{3}$\cite{MNEFGUI2014193}MAP  &   0.333(8) &  1.325(1)    \\
 $\gamma_{KLS}$ = -0.5 This work  &  0.343(3) &  1.267(7)     \\
% \hline
% La$_{0.7}$Ba$_{0.3}$MnO$_{3}$\cite{KHELIFI2014149}MAP  &   0.341(3) &  1.371(196)     \\
% La$_{0.7}$Ba$_{0.3}$MnO$_{3}$\cite{KHELIFI2014149}KF  &   0.357(160) &  1.341(4)      \\
% $q$ = 1.5 This work  &  0.343(3) &  1.267(7)     \\
 \hline
 Pr$_{0.77}$Pb$_{0.23}$MnO$_{3}$\cite{PhysRevB.75.024419}MAP  &   0.343(5) &  1.357(20)     \\
 Pr$_{0.77}$Pb$_{0.23}$MnO$_{3}$\cite{PhysRevB.75.024419}KF  &   0.344(1) &  1.352(6)     \\
 $\gamma_{KLS}$ = -0.30 This work  &  0.350(3) &  1.289(7)     \\
 \hline
 AMnO$_{3}$\cite{OMRI20123122}MAP  &   0.355(7) &  1.326(2)     \\
 AMnO$_{3}$\cite{OMRI20123122}KF  &   0.344(5) &  1.335(2)     \\
 $\gamma_{KLS}$ = -0.15 This work  &  0.357(3) &  1.309(7)     \\
 \hline
 Nd$_{0.7}$Pb$_{0.3}$MnO$_{3}$\cite{Ghosh_2005}MAP  &   0.361(13) &  1.325(1)           \\
 Nd$_{0.7}$Pb$_{0.3}$MnO$_{3}$\cite{Ghosh_2005}KF  &   0.361(5) &  1.314(1)           \\
 $\gamma_{KLS}$ = -0.10 This work  &  0.360(3) &  1.318(7)     \\
 \hline
 LaTi$_{0.2}$Mn$_{0.8}$O$_{3}$\cite{doi:10.1063/1.2795796}KF  &   0.359(4) &  1.280(10)     \\
 $\gamma_{KLS}$ = -0.10 This work  &  0.360(3) &  1.318(7)     \\
 \hline
 La$_{0.67}$Sr$_{0.33}$Mn$_{0.95}$V$_{0.05}$O$_{3}$\cite{MNEFGUI2014193}MAP  &   0.358(5) &  1.381(4)  \\
 $\gamma_{KLS}$ = -0.10 This work  &  0.360(3) &  1.318(7)     \\
 \hline
 \hline
 $\gamma_{KLS}$ = 0.00 \cite{GHODHBANE2013558} &  0.365(3) &  1.336(4)     \\
 \hline
 \hline
% La$_{0.67}$Sr$_{0.33}$Mn$_{0.95}$V$_{0.10}$O$_{3}$\cite{MNEFGUI2014193}MAP  &   0.367(3) &  1.414(2)  \\
% $q$ = 0.95 This work  &  0.369(3) &  1.347(7)     \\
% \hline
% La$_{0.67}$Ca$_{0.33}$MnO$_{3}$\cite{J.Appl.Phys.A.Berger}  &  0.368(3) &  1.384(17)  \\
% $q$ = 0.95 This work  &  0.369(3) &  1.347(7)     \\
% \hline
 Nd$_{0.85}$Pb$_{0.15}$MnO$_{3}$\cite{Ghosh_2005}MAP  &  0.372(1) &  1.340(30)  \\
 Nd$_{0.85}$Pb$_{0.15}$MnO$_{3}$\cite{Ghosh_2005}KF  &  0.372(4) &  1.347(1)  \\
 $\gamma_{KLS}$ = 0.05 This work  &  0.369(3) &  1.362(7)     \\
 \hline
 Nd$_{0.6}$Pb$_{0.4}$MnO$_{3}$\cite{PhysRevB.68.144408}KF  &   0.374(6) &  1.329(3)  \\
 $\gamma_{KLS}$ = 0.05 This work  &  0.369(3) &  1.362(7)     \\
 \hline
 La$_{0.67}$Sr$_{0.33}$Mn$_{0.95}$V$_{0.15}$O$_{3}$\cite{MNEFGUI2014193}MAP  &   0.375(3) &  1.355(6)  \\
 $\gamma_{KLS}$ = 0.05 This work  &  0.369(3) &  1.362(7)     \\
 \hline
 La$_{0.67}$Ba$_{0.22}$Sr$_{0.11}$MnO$_{3}$\cite{BenJemaa}MAP  &   0.378(3) &  1.388(1)   \\
 La$_{0.67}$Ba$_{0.22}$Sr$_{0.11}$MnO$_{3}$\cite{BenJemaa}KF  &   0.386(6) &  1.393(4)   \\
 $\gamma_{KLS}$ = 0.10 This work  &  0.377(3) &  1.347(7)     \\
 \hline
 LaTi$_{0.95}$Mn$_{0.05}$O$_{3}$\cite{doi:10.1063/1.2795796}KF  &   0.378(7) &  1.290(20)     \\
 $\gamma_{KLS}$ = 0.10 This work  &  0.377(3) &  1.347(7)     \\
 \hline
 LaTi$_{0.9}$Mn$_{0.1}$O$_{3}$\cite{doi:10.1063/1.2795796}KF  &   0.375(5) &  1.250(20)     \\
 $\gamma_{KLS}$ = 0.10 This work  &  0.377(3) &  1.347(7)     \\
 \hline
 LaTi$_{0.85}$Mn$_{0.15}$O$_{3}$\cite{doi:10.1063/1.2795796}KF  &   0.376(3) &  1.240(10)     \\
 $\gamma_{KLS}$ = 0.10 This work  &  0.377(3) &  1.347(7)     \\
 \hline
 La$_{0.67}$Ca$_{0.33}$Mn$_{0.9}$Ga$_{0.1}$O$_{3}$\cite{PhysRevB.70.104417}MAP  &  0.380(2) &  1.365(8)  \\
 La$_{0.67}$Ca$_{0.33}$Mn$_{0.9}$Ga$_{0.1}$O$_{3}$\cite{PhysRevB.70.104417}KF  &   0.387(6) &  1.362(2)  \\
 $\gamma_{KLS}$ = 0.15 This work  &  0.378(3) &  1.372(7)     \\
 \hline
 La$_{0.67}$Ba$_{0.22}$Sr$_{0.11}$Mn$_{0.9}$Fe$_{0.1}$O$_{3}$\cite{BenJemaa}MAP  &   0.398(2) &  1.251(5)   \\
 La$_{0.67}$Ba$_{0.22}$Sr$_{0.11}$Mn$_{0.9}$Fe$_{0.1}$O$_{3}$\cite{BenJemaa}KF  &   0.395(3) &  1.247(3)   \\
 $\gamma_{KLS}$ = 0.30 This work  &  0.395(3) &  1.424(7)     \\
 \hline
 La$_{0.67}$Ba$_{0.22}$Sr$_{0.11}$Mn$_{0.8}$Fe$_{0.2}$O$_{3}$\cite{BenJemaa}MAP  &   0.411(1) &  1.241(4)   \\
 La$_{0.67}$Ba$_{0.22}$Sr$_{0.11}$Mn$_{0.8}$Fe$_{0.2}$O$_{3}$\cite{BenJemaa}KF  &   0.394(3) &  1.292(3)   \\
 $\gamma_{KLS}$ = 0.30 This work  &  0.395(3) &  1.424(7)     \\
 \hline
 Pr$_{0.7}$Pb$_{0.3}$MnO$_{3}$\cite{PhysRevB.75.024419}MAP  &   0.404(6) &  1.354(20)     \\
 Pr$_{0.7}$Pb$_{0.3}$MnO$_{3}$\cite{PhysRevB.75.024419}KF  &   0.404(1) &  1.357(6)     \\
 $\gamma_{KLS}$ = 0.35 This work  &  0.402(3) &  1.446(7)     \\
 \hline
 La$_{0.75}$(Sr,Ca)$_{0.25}$Mn$_{0.9}$Ga$_{0.1}$O$_{3}$\cite{BenJemaa}MAP  &   0.420(5) &  1.221(2)   \\
 La$_{0.75}$(Sr,Ca)$_{0.25}$Mn$_{0.9}$Ga$_{0.1}$O$_{3}$\cite{BenJemaa}KF  &   0.428(5) &  1.286(4)   \\
 $\gamma_{KLS}$ = 0.45 This work  &  0.421(3) &  1.503(7)     \\
 \hline
 Pr$_{0.5}$Sr$_{0.5}$MnO$_{3}$\cite{PhysRevB.79.214426}MAP  &   0.443(2) &  1.339(6)     \\
 Pr$_{0.5}$Sr$_{0.5}$MnO$_{3}$\cite{PhysRevB.79.214426}KF  &   0.448(9) &  1.334(10)     \\
 $\gamma_{KLS}$ = 0.50 This work  &  0.434(3) &  1.540(7)     \\
 \hline
 \hline
\end{tabular}
\label{tableexponentsE3N3}
\end{table}
%\vspace{1cm}

\begin{table}[H]
\caption{Static $\gamma_{KLS}$-generalized critical exponents ($\gamma_{KLS} \neq 0$) to $3$d ($\epsilon = 1$) Ising ($N = 1$) systems, obtained from experiment through Modified Arrott (MA) plots \cite{PhysRevLett.19.786}, Kouvel-Fisher (KF) method \cite{PhysRev.136.A1626} and $\gamma_{KLS}$-SFT of this work.}
\begin{tabular}{ p{5.7cm}p{1.3cm}p{1.2cm}  }
 \hline
 %\multicolumn{4}{|c|}{Country List} \\
 \hline
 $\gamma_{KLS}$-Ising & $\beta_{\gamma_{KLS}}$ & $\gamma_{\gamma_{KLS}}$    \\
 \hline
 La$_{0.8}$Sr$_{0.2}$MnO$_{3}$\cite{J.Appl.Phys.Vasiliu-Doloca}  &   0.290(10) &        \\
 $\gamma_{KLS}$ = -0.35 This work  &  0.311(1) &  1.201(3)     \\
 \hline
 Nd$_{0.55}$Sr$_{0.45}$Mn$_{0.98}$Ga$_{0.02}$O$_{3}$\cite{YU2018393}  & 0.308(10) &  1.197 \\
 $\gamma_{KLS}$ = -0.35 This work  &  0.311(1) &  1.201(3)     \\
 \hline
 Pr$_{0.6}$Sr$_{0.4}$MnO$_{3}$\cite{PhysRevB.92.024409}MAP  &   0.315(0) &  1.095(7)      \\
 Pr$_{0.6}$Sr$_{0.4}$MnO$_{3}$\cite{PhysRevB.92.024409}KF  &   0.312(2) &  1.106(5)      \\
 $\gamma_{KLS}$ = -0.15 This work  &  0.318(1) &  1.221(3)     \\
 \hline
 La$_{0.8}$Ca$_{0.2}$MnO$_{3}$\cite{ZHANG2013146}KF  &   0.316(7) &  1.081(36)      \\
 $\gamma_{KLS}$ = -0.15 This work  &  0.318(1) &  1.221(3)     \\
 \hline
 La$_{0.7}$Ca$_{0.3}$Mn$_{0.85}$Ni$_{0.15}$O$_{3}$\cite{PHAN201440}MAP  &   0.320(9) &  0.990(82)      \\
 $\gamma_{KLS}$ = -0.05 This work  &  0.322(1) &  1.234(3)     \\
 \hline
 Nd$_{0.6}$Sr$_{0.4}$MnO$_{3}$\cite{RSCAdvJeddi}MAP  &   0.320(6) &  1.239(2)      \\
 Nd$_{0.6}$Sr$_{0.4}$MnO$_{3}$\cite{RSCAdvJeddi}KF  &   0.323(2) &  1.235(4)      \\
 $\gamma_{KLS}$ = -0.05 This work  &  0.322(1) &  1.234(3)     \\
 \hline
 Nd$_{0.6}$Sr$_{0.4}$MnO$_{3}$\cite{PhysRevB.92.024409}MAP  &   0.321(3) &  1.183(17)      \\
 Nd$_{0.6}$Sr$_{0.4}$MnO$_{3}$\cite{PhysRevB.92.024409}KF  &   0.308(4) &  1.172(11)      \\
 $\gamma_{KLS}$ = -0.05 This work  &  0.322(1) &  1.234(3)     \\
 \hline
 Nd$_{0.67}$Ba$_{0.33}$MnO$_{3}$\cite{HCINI20152042}MAP  &   0.325(4) &  1.248(19)      \\
 Nd$_{0.67}$Ba$_{0.33}$MnO$_{3}$\cite{HCINI20152042}KF  &   0.326(5) &  1.244(33)      \\
 $\gamma_{KLS}$ = -0.05 This work  &  0.322(1) &  1.234(3)     \\
 \hline
 $\gamma_{KLS}$ = 0.00 \cite{GHODHBANE2013558} &  0.325(2) &  1.241(2)     \\
 \hline
 La$_{0.65}$Bi$_{0.05}$Sr$_{0.3}$MnO$_{3}$\cite{Phys.SolidStateBaazaoui}MAP  &   0.335(3) &  1.207(20)      \\
 La$_{0.65}$Bi$_{0.05}$Sr$_{0.3}$MnO$_{3}$\cite{Phys.SolidStateBaazaoui}KF  &   0.316(7) &  1.164(20)      \\
 $\gamma_{KLS}$ = 0.10 This work  &  0.330(1) &  1.258(3)     \\
 \hline
 La$_{0.65}$Bi$_{0.05}$Sr$_{0.3}$Mn$_{0.94}$Ga$_{0.06}$O$_{3}$\cite{Phys.SolidStateBaazaoui}MAP & 0.334(4) & 1.192(8) \\
 La$_{0.65}$Bi$_{0.05}$Sr$_{0.3}$Mn$_{0.94}$Ga$_{0.06}$O$_{3}$\cite{Phys.SolidStateBaazaoui}KF  & 0.307(8) & 1.138(5) \\
 $\gamma_{KLS}$ = 0.10 This work  &  0.330(1) &  1.258(3)     \\
 \hline
 \hline
 \end{tabular}
\label{tableexponentsE3N1}
\end{table}
%\vspace{1cm}

\begin{table}[t]
\caption{$\gamma_{KLS}$-generalized critical temperature ($\gamma_{KLS} \neq 0$) to $3$d ($\epsilon = 1$) Heisenberg ($N = 3$) systems, obtained from experiment through Modified Arrott (MA) plots \cite{PhysRevLett.19.786} and Kouvel-Fisher (KF) methods.}
\begin{tabular}{ p{5.7cm}p{1.4cm}p{.9cm}  }
 \hline
 %\multicolumn{4}{|c|}{Country List} \\
 \hline
 $\gamma_{KLS}$-Heisenberg & $\beta_{\gamma_{KLS}}$ & $T_{c,\hspace{.5mm}\gamma_{KLS}}(K)$    \\
 \hline
 La$_{0.67}$Sr$_{0.33}$MnO$_{3}$\cite{MNEFGUI2014193}MAP  &   0.333(8) &  350    \\ 
 $\gamma_{KLS}$ = -0.5 This work  &  0.343(3) &       \\
 \hline
 La$_{0.67}$Sr$_{0.33}$Mn$_{0.95}$V$_{0.05}$O$_{3}$\cite{MNEFGUI2014193}MAP  &   0.358(5) &  326  \\
 $\gamma_{KLS}$ = -0.10 This work  &  0.360(3) &       \\
 \hline
 La$_{0.67}$Sr$_{0.33}$Mn$_{0.95}$V$_{0.10}$O$_{3}$\cite{MNEFGUI2014193}MAP  &   0.367(3) &  301  \\
 $\gamma_{KLS}$ = 0.05 This work  &  0.369(3) &       \\
 \hline
 La$_{0.67}$Sr$_{0.33}$Mn$_{0.95}$V$_{0.15}$O$_{3}$\cite{MNEFGUI2014193}MAP  &   0.375(3) &  290  \\
 $\gamma_{KLS}$ = 0.10 This work  &  0.377(3) &       \\
 \hline
 \hline
 La$_{0.67}$Ba$_{0.22}$Sr$_{0.11}$MnO$_{3}$\cite{BenJemaa}MAP  &   0.378(3) &  343   \\
 La$_{0.67}$Ba$_{0.22}$Sr$_{0.11}$MnO$_{3}$\cite{BenJemaa}KF  &   0.386(6) &   342   \\
 $\gamma_{KLS}$ = 0.10 This work  &  0.377(3) &       \\
 \hline
 La$_{0.67}$Ba$_{0.22}$Sr$_{0.11}$Mn$_{0.9}$Fe$_{0.1}$O$_{3}$\cite{BenJemaa}MAP  &   0.398(2) &  191   \\
 La$_{0.67}$Ba$_{0.22}$Sr$_{0.11}$Mn$_{0.9}$Fe$_{0.1}$O$_{3}$\cite{BenJemaa}KF   &   0.395(3) &  189   \\
 $\gamma_{KLS}$ = 0.30 This work  &  0.395(3) &       \\
 \hline
 La$_{0.67}$Ba$_{0.22}$Sr$_{0.11}$Mn$_{0.8}$Fe$_{0.2}$O$_{3}$\cite{BenJemaa}MAP  &   0.411(1) &  130   \\
 La$_{0.67}$Ba$_{0.22}$Sr$_{0.11}$Mn$_{0.8}$Fe$_{0.2}$O$_{3}$\cite{BenJemaa}KF   &   0.394(3) &  139   \\ 
 $\gamma_{KLS}$ = 0.30 This work  &  0.395(3) &       \\
 \hline
 \hline
 La$_{0.7}$Ba$_{0.3}$MnO$_{3}$\cite{KHELIFI2014149}MAP  &   0.341(3)  &  339     \\
 La$_{0.7}$Ba$_{0.3}$MnO$_{3}$\cite{KHELIFI2014149}KF  &   0.357(160) &  340      \\
 $\gamma_{KLS}$ = -0.50 This work  &  0.343(3) &      \\
 \hline
 La$_{0.6}$Pr$_{0.1}$Ba$_{0.3}$MnO$_{3}$\cite{KHELIFI2014149}MAP  &   0.396(16)   &  321     \\
 La$_{0.6}$Pr$_{0.1}$Ba$_{0.3}$MnO$_{3}$\cite{KHELIFI2014149}KF   &   0.391(2)    &  321      \\
 $\gamma_{KLS}$ = 0.30 This work  &  0.395(3) &       \\
 \hline
 La$_{0.5}$Pr$_{0.2}$Ba$_{0.3}$MnO$_{3}$\cite{KHELIFI2014149}MAP  &   0.494(10)   &  304     \\
 La$_{0.5}$Pr$_{0.2}$Ba$_{0.3}$MnO$_{3}$\cite{KHELIFI2014149}KF   &   0.491(23)   &  304      \\
 $\gamma_{KLS}$ = 0.50 This work  &  0.434(3) &       \\
 \hline
 \hline
 \end{tabular}
\label{tablecriticalTC}
\end{table}

\begin{table}[t]
\caption{Static critical exponents to $3$d ($\epsilon = 1$) Heinsenberg ($N = 3$) systems, obtained from experiment through Modified Arrott (MA) plots \cite{PhysRevLett.19.786}, Kouvel-Fisher (KF) method \cite{PhysRev.136.A1626} and NSFT ($q \neq 0$) of Ref. \cite{CARVALHO2023137683}.}
\begin{tabular}{ p{5.0cm}p{1.4cm}p{1.4cm}  }
 \hline
 %\multicolumn{4}{|c|}{Country List} \\
 \hline
 Heisenberg & $\beta$ & $\gamma$    \\
 \hline
 La$_{0.7}$Sr$_{0.3}$MnO$_{0.97}$Ni$_{0.03}$O$_{3}$\cite{GINTING201317}  &   0.468(6) &  1.010(21)     \\
 $\gamma_{KLS}$ = - This work  &  - &  -     \\
 $q$ = 0.40 \cite{CARVALHO2023137683}  &  0.469(4) &  1.640(8)     \\
 \hline
 La$_{0.7}$Sr$_{0.3}$Mn$_{0.94}$Co$_{0.06}$O$_{3}$\cite{PHAN2014S247}  &   0.478(13) &  1.165(27)     \\
 $\gamma_{KLS}$ = - This work  &  - &  -     \\
 $q$ = 0.39 \cite{CARVALHO2023137683}  &  0.474(4) &  1.653(8)     \\
 \hline
 La$_{0.7}$Sr$_{0.3}$Mn$_{0.92}$Co$_{0.08}$O$_{3}$\cite{PHAN2014S247}  &   0.483(18) &  1.112(28)     \\
 $\gamma_{KLS}$ = - This work  &  - &  -     \\
 $q$ = 0.37 \cite{CARVALHO2023137683}  &  0.484(4) &  1.680(8)     \\
 \hline
 La$_{0.7}$Sr$_{0.3}$Mn$_{0.90}$Co$_{0.10}$O$_{3}$\cite{PHAN2014S247}  &   0.487(16) &  1.109(63)     \\
 $\gamma_{KLS}$ = - This work  &  - &  -     \\
 $q$ = 0.36 \cite{CARVALHO2023137683}  &  0.489(4) &  1.695(8)     \\
 \hline
 La$_{0.67}$Ca$_{0.33}$Mn$_{0.95}$Fe$_{0.05}$O$_{3}$\cite{NISHA201266}  &   0.550(10) &  1.0246(3)     \\
 $\gamma_{KLS}$ = - This work  &  - &  -     \\
 $q$ = 0.27 \cite{CARVALHO2023137683}  &  0.556(4) &  1.876(9)     \\
 \hline
 La$_{0.67}$Ca$_{0.33}$Mn$_{0.90}$Cr$_{0.10}$O$_{3}$\cite{NISHA201240}  &   0.555(6) &   1.170(40)     \\
 $\gamma_{KLS}$ = - This work  &  - &  -     \\
 $q$ = 0.27 \cite{CARVALHO2023137683}  &  0.556(4) &  1.876(9)     \\
 \hline
 La$_{0.67}$Ca$_{0.33}$Mn$_{0.75}$Cr$_{0.25}$O$_{3}$\cite{NISHA201240}  &  0.680(10) &  1.090(30)     \\
 $\gamma_{KLS}$ = - This work  &  - &  -     \\
 $q$ = 0.19 \cite{CARVALHO2023137683}  &  0.674(5) &  2.172(10)     \\
 \hline
 \hline
 \end{tabular}
\label{tableexponentsNSFT}
\end{table}
%\vspace{1cm}

%\section{Physical interpretation}

\par We interpret these results as follows: a given physical quantity, near the transition point, diverges. How much it diverges is measured by its associated critical exponent. In the case of susceptibility, for example, its inverse furnishes a measure of how much the material is susceptible to the changes in magnetic field. Higher (lower) values of the $\gamma$ critical index indicates more (less) susceptible or weaker interacting (stronger) systems. These facts are in agreement with the form of the effective energy of the system, which can be obtained by the some expansion around $\gamma_{KLS} \approx 0$. In fact, taking the leading contribution to the energy of the system as $E$, in units of $k_{B}T$, from $e_{\gamma_{KLS}}^{-E} \approx e^{-E}\left(1 - \frac{1}{2}\gamma_{KLS} E^{2}\right) \approx e^{-\left(E + \frac{1}{2}\gamma_{KLS} E^{2}\right)}$. So the effective energy $E + \frac{1}{2}\gamma_{KLS} E^{2}$ increases with the increasing of $\gamma_{KLS}$. Then higher (lower) values of $\gamma_{KLS}$ represent systems interacting weaker (stronger) or more (less) susceptible and thus possessing higher (lower) values of their critical exponents. Also higher (lower) values of $\gamma_{KLS}$ give higher (lower) values of $E$ and then we have to furnish less (more) energy to attain the respective critical transition temperature so the critical transition temperatures assume lower (higher) values and decrease. Although $\gamma_{KLS}$-\textit{SFT} can explain the results for the Tables \ref{tableexponentsE3N3}-\ref{tableexponentsE3N1}, it does not explain the results of Table  \ref{tableexponentsNSFT} for some materiais, being explained only for NSFT (see Table \ref{tableexponentsNSFT}) of Ref. \cite{CARVALHO2023137683}. Then, $\gamma_{KLS}$-\textit{SFT} must be discarded as one trying to generalize statistical mechanics. For our knowledge, only NSFT of Ref. \cite{CARVALHO2023137683} remains a fully consistent or complete generalized formulation of statistical field theory.

\section{Other $\gamma_{KLS}$-models}

\par Now that $\gamma_{KLS}$-SFT has been validated experimentally, we display the $\gamma_{KLS}$-critical indices for other models.

\subsection{Both $\gamma_{KLS}$-percolation and $\gamma_{KLS}$-Yang-lee edge singularity}

\par In the case of both $\gamma_{KLS}$-percolation \cite{Bonfirm_1981} ($\alpha = -1$ and $\beta = -2$) and $\gamma_{KLS}$-Yang-Lee edge singularity \cite{Bonfirm_1981} ($\alpha = -1$ and $\beta = -1$) in dimensions $d = 6 - \epsilon$ we can write, 
\begin{eqnarray}
&&\eta_{\gamma_{KLS}} = \eta - \nonumber \\&& \frac{\gamma_{KLS}}{1 - \gamma_{KLS}}\frac{8\alpha\beta}{3(\alpha - 4\beta)[\alpha - 4\beta(1 - 3\gamma_{KLS})/(1 - \gamma_{KLS})]}\epsilon ,\nonumber \\&&
\end{eqnarray}
\begin{eqnarray}
&&\nu_{\gamma_{KLS}}^{-1} = \nu^{-1} - \nonumber \\&&\frac{\gamma_{KLS}}{1 - \gamma_{KLS}}\frac{40\alpha\beta}{3(\alpha - 4\beta)[\alpha - 4\beta(1 - 3\gamma_{KLS})/(1 - \gamma_{KLS})]}\epsilon ,\nonumber \\&&
\end{eqnarray}
\begin{eqnarray}
\omega_{\gamma_{KLS}} = \omega .
\end{eqnarray}

%\begin{eqnarray}\label{etaphi3}
%\eta_{\gamma} = \eta + (q - 1)\frac{8\alpha\beta}{3(\alpha - 4\beta)[\alpha - 4\beta(2q - 1)]}\epsilon , 
%\end{eqnarray}
%\begin{eqnarray}\label{nuphi3}
%\nu_{\gamma}^{-1} = \nu^{-1} + (q - 1)\frac{40\alpha\beta}{3(\alpha - 4\beta)[\alpha - 4\beta(2q - 1)]}\epsilon ,
%\end{eqnarray}
%\begin{eqnarray}\label{omegaphi3}
%\omega_{\gamma} = \omega .
%\end{eqnarray}
For $\gamma_{KLS}$-Yang-Lee edge singularity, the $\eta_{\gamma_{KLS}}$ and $\nu_{\gamma_{KLS}}$ indices are dependent \cite{Bonfirm_1981,PhysRevD.95.085001}. They are related through $\nu_{\gamma_{KLS}}^{-1} = (d - 2 + \eta_{\gamma_{KLS}})/2$ \cite{Bonfirm_1981,PhysRevD.95.085001}. Thus from the value of $\eta_{\gamma_{KLS}}$, by employing the values $\alpha = -1$ and $\beta = -1$, we compute $\nu_{\gamma_{KLS}}$. The remaining ones can be obtained from the scaling relations among them \cite{PhysRevD.103.116024}.

\subsection{$\gamma_{KLS}$-$\phi^{6}$ theory}

\par The $N$-component $\gamma_{KLS}$-$\phi^{6}$ theory in $d = 3 - \epsilon$ \cite{STEPHEN197389,PhysRevE.60.2071} describes $\gamma_{KLS}$-tricritical points to $N = 2$ and corresponds to $^{3}$He-$^{4}$He mixtures or antiferromagnets in the presence of a strong external field (metamagnets) \cite{Hager_2002}. The $\gamma_{KLS}$-critical exponents assume the form 
%$\eta_{\gamma} = \eta + \frac{\gamma}{1 - \gamma}\frac{(N + 2)(N + 4)}{12(3N + 22)^{2}}\epsilon^{2}$, $\nu_{\gamma} = \nu + \frac{\gamma}{1 - \gamma}\frac{(N + 2)(N + 4)}{3(3N + 22)^{2}}\epsilon^{2}$.
\begin{eqnarray}
\eta_{\gamma_{KLS}} = \eta + \frac{\gamma_{KLS}}{1 - \gamma_{KLS}}\frac{(N + 2)(N + 4)}{12(3N + 22)^{2}}\epsilon^{2}, 
\end{eqnarray}
\begin{eqnarray}
\nu_{\gamma_{KLS}} = \nu + \frac{\gamma_{KLS}}{1 - \gamma_{KLS}}\frac{(N + 2)(N + 4)}{3(3N + 22)^{2}}\epsilon^{2}.
\end{eqnarray}
Actually, $\eta$ and $\nu$ are known up to six-loop order \cite{Hager_2002}.

\subsection{$\gamma_{KLS}$-long-range systems}

\par The $\gamma_{KLS}$-long-range critical exponents for $N$-component $\gamma_{KLS}$-Ising-like models whit interactions decaying as $1/r_{ij}^{d + \sigma}$ can be defined at three different sectors \cite{BrezinEandParisiGandRicci-TersenghiF}. The $\gamma_{KLS}$-critical indices, in $d = 2\sigma - \varepsilon$, are given by 
%$\eta_{\sigma, \hspace{.5mm}\gamma} = \eta_{\gamma}, \eta_{\sigma}, 0$ and $\nu_{\sigma, \hspace{.5mm}\gamma} = \nu_{\gamma}, \nu_{\sigma}, \nu_{\sigma} + \frac{\gamma}{1 - \gamma}\frac{(N + 2)}{\sigma^{2}(N + 8)}\varepsilon$ for $\sigma > 2 - \eta_{\gamma}$, $d/2 < \sigma < 2 - \eta_{\gamma}$, $\sigma < d/2$, respectively.   
%\[   
%\eta_{\sigma, \hspace{.5mm}\gamma},\hspace{.5mm} \nu_{\sigma, \hspace{.5mm}\gamma} = 
%     \begin{cases}
%       \eta_{\gamma},\hspace{.5mm}\nu_{\gamma},\hspace{1mm} \sigma > 2 - \eta_{\gamma} \\
%       \eta_{\sigma},\hspace{.5mm} \nu_{\sigma} + \frac{\gamma}{1 - \gamma}\frac{(N + 2)}{\sigma^{2}(N + 8)}\varepsilon ,\hspace{1mm} d/2 < \sigma < 2 - \eta_{\gamma} \\
%       0,\hspace{.5mm}1/2,\hspace{1mm} \sigma < d/2 , \\
%     \end{cases}
%\]
\[
\eta_{\sigma, \hspace{.5mm}\gamma_{KLS}} = 
     \begin{cases}
       \eta_{\gamma_{KLS}} &\quad\text{if}\quad \sigma > 2 - \eta_{\gamma_{KLS}} \\
       \eta_{\sigma}  &\quad\text{if}\quad d/2 < \sigma < 2 - \eta_{\gamma_{KLS}} \\
       0 &\quad\text{if}\quad \sigma < d/2 , \\
     \end{cases}
\]
\[
\nu_{\sigma, \hspace{.5mm}\gamma_{KLS}} = 
     \begin{cases}
       \nu_{\gamma_{KLS}} &\quad\text{if}\quad \sigma > 2 - \eta_{\gamma_{KLS}} \\
       \nu_{\sigma} + \frac{\gamma_{KLS}}{1 - \gamma_{KLS}}\frac{(N + 2)}{\sigma^{2}(N + 8)}\epsilon &\quad\text{if}\quad d/2 < \sigma < 2 - \\\eta_{\gamma_{KLS}} \\
       1/2 &\quad\text{if}\quad \sigma < d/2 , \\
     \end{cases}
\]
where $\eta_{\sigma} = 2 - \sigma$ for any loop order \cite{LohmannMSladeGLallaceBC,SladeG}, $\eta_{\gamma_{KLS}}$ and $\nu_{\gamma_{KLS}}$ are the short-range $\gamma_{KLS}$-critical indices of Eqs. (\ref{etaphi4})-(\ref{nuphi4}), and $\nu_{\sigma}$ has been evaluated up to two-loop order \cite{PhysRevLett.29.917} in the interval $d/2 < \sigma < 2 - \eta_{\gamma_{KLS}}$.

%$\mathcal{H} =  -\sum_{<ij>}\frac{J}{r_{ij}^{d + \sigma}}S_{i}S_{j}$

\subsection{$\gamma_{KLS}$-Gross-Neveu model}

\par From the $\gamma_{KLS}$-Gross-Neveu model in $d = 2 + \epsilon$ \cite{PhysRevD.10.3235} we can study both $\psi$ and $\bar{\psi}$ Dirac massive fermions of mass $\mathcal{M}$. The $\gamma_{KLS}$-critial exponents are 
%$\eta_{\psi, \hspace{.5mm}\gamma} = \eta_{\psi} + \frac{\gamma}{1 - \gamma}\frac{(2N - 1)}{8(N - 1)^{2}}\epsilon^{2}$, $\eta_{\mathcal{M}, \hspace{.5mm}\gamma} = \eta_{\mathcal{M}} + \frac{\gamma}{1 - \gamma}\frac{\epsilon}{2(N - 1)}$, $\nu_{\gamma} = \nu$.
\begin{eqnarray}
\eta_{\psi, \hspace{.5mm}\gamma_{KLS}} = \eta_{\psi} + \frac{\gamma_{KLS}}{1 - \gamma_{KLS}}\frac{(2N - 1)}{8(N - 1)^{2}}\epsilon^{2}, 
\end{eqnarray}
\begin{eqnarray}
\eta_{\mathcal{M}, \hspace{.5mm}\gamma_{KLS}} = \eta_{\mathcal{M}} + \frac{\gamma_{KLS}}{1 - \gamma_{KLS}}\frac{\epsilon}{2(N - 1)}, \hspace{.3cm} \nu_{\gamma_{KLS}} = \nu ,
\end{eqnarray}
%\begin{eqnarray}
%\nu_{\gamma_{KLS}} = \nu ,
%\end{eqnarray}
where, up to now, the nongeneralized critical exponents have been evaluated up to four-loop level \cite{PhysRevD.94.125028}.

\subsection{$\gamma_{KLS}$-uniaxial systems with strong dipolar forces}

\par $\gamma_{KLS}$-uniaxial systems with strong dipolar forces in the $z$-direction are described by the following Hamiltonian \cite{PhysRevB.13.251} 
%\\ $\mathcal{H} =  -\sum_{\vec{x}\vec{x}^{\prime}}\sum_{\mu\nu}S_{\vec{x}}^{\mu}S_{\vec{x}^{\prime}}^{\nu}\left(V_{\mu\nu}(\vec{x} - \vec{x}^{\prime}) +\frac{ \gamma\partial_{z}\partial_{z^{\prime}}}{|\vec{x} - \vec{x}^{\prime}|^{d - 2}}\right)$,
\begin{eqnarray}
\mathcal{H} =  -\sum_{\vec{x}\vec{x}^{\prime}}\sum_{\mu\nu}S_{\vec{x}}^{\mu}S_{\vec{x}^{\prime}}^{\nu}\left(V_{\mu\nu}(\vec{x} - \vec{x}^{\prime}) +\frac{ \gamma\partial_{z}\partial_{z^{\prime}}}{|\vec{x} - \vec{x}^{\prime}|^{d - 2}}\right),\hspace{5mm} 
\end{eqnarray}
where $V_{\mu\nu}(\vec{x})$ is the short-range potential and $\gamma$ is a parameter for controlling the dipolar forces intensity. In in $d = 3 - \epsilon$ dimensions, the $\gamma_{KLS}$-critical exponents are given by 
%$\eta_{\gamma} = \eta + \frac{\gamma}{1 - \gamma}\frac{4(N + 2)}{9(N + 8)^{2}}\epsilon^{2}$, $\nu_{\gamma}^{-1} = \nu^{-1} - \frac{\gamma}{1 - \gamma}\frac{(N + 2)}{(N + 8)}\epsilon$.
\begin{eqnarray}
\eta_{\gamma_{KLS}} = \eta + \frac{\gamma_{KLS}}{1 - \gamma_{KLS}}\frac{4(N + 2)}{9(N + 8)^{2}}\epsilon^{2},  
\end{eqnarray}
\begin{eqnarray}
\nu_{\gamma_{KLS}}^{-1} = \nu^{-1} - \frac{\gamma_{KLS}}{1 - \gamma_{KLS}}\frac{(N + 2)}{(N + 8)}\epsilon ,
\end{eqnarray}
where, $\eta$ and $\nu$ are displayed in Ref. \cite{PhysRevB.13.251} up to two-loop level.

\subsection{$\gamma_{KLS}$-spherical model}

\par The $\gamma_{KLS}$-spherical model \cite{PhysRev.86.821} can be obtained by taking the limit $N\rightarrow\infty$ \cite{PhysRev.176.718} of the O($N$)$_{\gamma_{KLS}}$ model of the present Letter. After taking this limit, we obtain in $d = 4 - \epsilon$ 
%$\eta_{\gamma} = \eta$, $\nu_{\gamma} = \nu + \frac{\gamma}{1 - \gamma}\frac{\epsilon}{4}$. 
\begin{eqnarray}
\eta_{\gamma_{KLS}} = \eta ,  \hspace{2cm} \nu_{\gamma_{KLS}} = \nu + \frac{\gamma_{KLS}}{1 - \gamma_{KLS}}\frac{\epsilon}{4},
\end{eqnarray}
%\begin{eqnarray}
%\nu_{\gamma_{KLS}} = \nu + \frac{\gamma_{KLS}}{1 - \gamma_{KLS}}\frac{\epsilon}{4},
%\end{eqnarray}
where the corresponding nongeneralized values are exact, namely $\eta = 0$ and $\nu = 1/(2 - \epsilon)$ \cite{PhysRevLett.28.240}.

\subsection{$\gamma_{KLS}$-Lifshitz critical points}

\par The $\gamma_{KLS}$-Lifshitz points \cite{PhysRevLett.35.1678,PhysRevB.67.104415,PhysRevB.72.224432,Albuquerque_2001,LEITE2004281,PhysRevB.61.14691,PhysRevB.68.052408,FARIAS,Borba,Santos_2014,deSena_2015,Santos_2019,Santos_20192,Leite_2022} are composed of the $m$-axial Lifshitz points \cite{PhysRevB.67.104415} an their generalized forms for the higher character cases \cite{PhysRevB.72.224432}. For the latter, there are $d - m$-, $m_{2}$-,..., $m_{n}$-dimensional vectors, respectively. The generic higher character $\gamma_{KLS}$-Lifshitz anisotropic (treated both in the orthogonal approximation) and isotropic (computed in both orthogonal approximation and exactly) indices are given by  
%$\eta_{n, \hspace{.5mm}\gamma} = \eta_{n} + \frac{\gamma}{1 - \gamma}n\frac{(N + 2)}{2(N + 8)^{2}}\epsilon_{n}^{2}$, $\nu_{n, \hspace{.5mm}\gamma} = \nu_{n} + \frac{\gamma}{1 - \gamma}\frac{(N + 2)}{4n(N + 8)}\epsilon_{n}$,
\begin{eqnarray}
\eta_{n, \hspace{.5mm}\gamma_{KLS}} = \eta_{n} + \frac{\gamma_{KLS}}{1 - \gamma_{KLS}}n\frac{(N + 2)}{2(N + 8)^{2}}\epsilon_{n}^{2}, 
\end{eqnarray}
\begin{eqnarray}
\nu_{n, \hspace{.5mm}\gamma_{KLS}} = \nu_{n} + \frac{\gamma_{KLS}}{1 - \gamma_{KLS}}\frac{(N + 2)}{4n(N + 8)}\epsilon_{n},
\end{eqnarray}
where $\epsilon_{L} = 4 + \sum_{n = 2}^{L}[(n - 1)/n]m_{n} - d$, 
%$\eta_{n, \hspace{.5mm}\gamma} = \eta_{n} + \frac{\gamma}{1 - \gamma}\frac{(N + 2)}{2n(N + 8)^{2}}\epsilon_{n}^{2}$, $\nu_{n, \hspace{.5mm}\gamma} = \nu_{n} + \frac{\gamma}{1 - \gamma}\frac{(N + 2)}{4n^{2}(N + 8)}\epsilon_{n}$, 
\begin{eqnarray}
\eta_{n, \hspace{.5mm}\gamma_{KLS}} = \eta_{n} + \frac{\gamma_{KLS}}{1 - \gamma_{KLS}}\frac{(N + 2)}{2n(N + 8)^{2}}\epsilon_{n}^{2}, 
\end{eqnarray}
\begin{eqnarray}
\nu_{n, \hspace{.5mm}\gamma_{KLS}} = \nu_{n} + \frac{\gamma_{KLS}}{1 - \gamma}\frac{(N + 2)}{4n^{2}(N + 8)}\epsilon_{n},
\end{eqnarray}
where $\epsilon_{L} = 4n - d$, 
%$\eta_{n, \hspace{.5mm}\gamma} = \eta_{n} + \frac{\gamma}{1 - \gamma}\frac{(-1)^{n + 1}\gamma(2n)^{2}(N + 2)}{\gamma(n + 1)\gamma(3n)(N + 8)^{2}}\epsilon_{n}^{2}$, $\nu_{n, \hspace{.5mm}\gamma} = \nu_{n} + \frac{\gamma}{1 - \gamma}\frac{(N + 2)}{4n^{2}(N + 8)}\epsilon_{n}$, respectively.
\begin{eqnarray}
\eta_{n, \hspace{.5mm}\gamma} = \eta_{n} + \frac{\gamma}{1 - \gamma}\frac{(-1)^{n + 1}\Gamma(2n)^{2}(N + 2)}{\Gamma(n + 1)\Gamma(3n)(N + 8)^{2}}\epsilon_{n}^{2}, 
\end{eqnarray}
\begin{eqnarray}
\nu_{n, \hspace{.5mm}\gamma} = \nu_{n} + \frac{\gamma}{1 - \gamma}\frac{(N + 2)}{4n^{2}(N + 8)}\epsilon_{n}.
\end{eqnarray}
In Ref. \cite{PhysRevB.72.224432}, the $\gamma_{KLS}$-critical indices were obtained up to next-to-leading order.

% demais referências \cite{PhysRevLett.35.1678,PhysRevB.67.104415,PhysRevB.72.224432,Albuquerque_2001,LEITE2004281,PhysRevB.61.14691,PhysRevB.68.052408,FARIAS,Borba,Santos_2014,deSena_2015,Santos_2019,Santos_20192,Leite_2022}

\subsection{Long-range $\gamma_{KLS}$-$\lambda\phi^{3}$ theory}

\par The long-range $\gamma_{KLS}$-$\lambda\phi^{3}$ \cite{PhysRevB.31.379} critical indices in $d = 3\sigma - \varepsilon$ are given by 
%$\eta_{\sigma , \hspace{.5mm}\gamma} = \eta_{\sigma}$, $\nu_{\sigma, \hspace{.5mm}\gamma}^{-1} = \nu_{\sigma}^{-1} - \frac{2\gamma}{1 - 3\gamma}\frac{\alpha}{\beta}\varepsilon$.
\begin{eqnarray}
\eta_{\sigma , \hspace{.5mm}\gamma_{KLS}} = \eta_{\sigma}, \hspace{.7cm} \nu_{\sigma , \hspace{.5mm}\gamma_{KLS}}^{-1} = \nu_{\sigma}^{-1} - \frac{\gamma_{KLS}}{1 - 3\gamma_{KLS}}\frac{\alpha}{\beta}\epsilon ,
\end{eqnarray}
%\begin{eqnarray}
%\nu_{\sigma , \hspace{.5mm}\gamma_{KLS}}^{-1} = \nu_{\sigma}^{-1} - \frac{\gamma_{KLS}}{1 - 3\gamma_{KLS}}\frac{\alpha}{\beta}\epsilon ,
%\end{eqnarray}
The values of the constants $\alpha$ and $\beta$ are $-1$, $-1$ and $-1$, $-2$ for the nongeneralized Yang-Lee edge singularity and percolation \cite{Bonfirm_1981}, respectively. The nongeneralized critical exponent $\eta_{\sigma} = 2 - \sigma$ is exact \cite{PhysRevB.31.379}. Then the index $\eta_{\sigma ,\hspace{.5mm}\gamma_{KLS}}$ is exact. nongeneralized exponents were obtained up to two-loop level in Ref. \cite{PhysRevB.31.379}.

\subsection{$\gamma_{KLS}$-multicritical points}

\par The $\gamma_{KLS}$-multicritical points of order $k$ \cite{C.ItzyksonJ.M.Drouffe} whose critical dimension is $d_{c} = \frac{2k}{k - 1}$ possess the following $\gamma_{KLS}$-critical exponents, namely  
%$\eta_{\gamma} = \eta + \frac{\gamma}{1 - \gamma}4(k - 1)^{2}\frac{(k)!^{6}}{(2k!)^{3}}\epsilon^{2}$
\begin{eqnarray}
\eta_{\gamma_{KLS}} = \eta + \frac{\gamma_{KLS}}{1 - \gamma_{KLS}}4(k - 1)^{2}\frac{(k)!^{6}}{(2k!)^{3}}\epsilon^{2}.
\end{eqnarray}
The nongeneralized index $\eta$ was obtained in Ref. \cite{C.ItzyksonJ.M.Drouffe} up to $2k - 2$-loop level.

\subsection{$\gamma_{KLS}$-Gross-Neveu-Yukawa model}

\par With the $\gamma_{KLS}$-Gross-Neveu-Yukawa model \cite{ZINNJUSTIN1991105}, in $d = 4 - \epsilon$, we can study the interaction between one scalar field $\phi$ and $N$ massless Dirac fermions $\psi$ and $\bar{\psi}$. The $\gamma_{KLS}$-critical indices are given by  
%$\eta_{\psi ,\hspace{.5mm}\gamma} = \eta_{\psi} + \gamma\frac{2\epsilon}{(2N + 3)[2N + 3 - (2N + 7)\gamma]}$, $\eta_{\phi ,\hspace{.5mm}\gamma} = \eta_{\phi} + \gamma\frac{8N\epsilon}{(2N + 3)[2N + 3 - (2N + 7)\gamma]}$, $\nu_{\gamma}^{-1} = \nu^{-1} - \frac{A_{N,\hspace{.5mm}q}}{(2N + 3)[2N + 1 + 2(2q - 1)]}\epsilon$,   
\begin{eqnarray}
&&\eta_{\psi ,\hspace{.5mm}\gamma_{KLS}} = \eta_{\psi} + \nonumber \\&&\gamma_{KLS}\frac{2\epsilon}{(2N + 3)[(2N + 1)(1 - \gamma_{KLS}) + 2(1 - 3\gamma_{KLS})]} ,\nonumber \\&& 
\end{eqnarray}
\begin{eqnarray}
&&\eta_{\phi ,\hspace{.5mm}\gamma_{KLS}} = \eta_{\phi} + \nonumber \\&&\gamma_{KLS}\frac{8N\epsilon}{(2N + 3)[(2N + 1)(1 - \gamma_{KLS}) + 2(1 - 3\gamma_{KLS})]} ,\nonumber \\&&
\end{eqnarray}
\begin{eqnarray}
&&\nu_{\gamma_{KLS}}^{-1} = \nu^{-1} - \nonumber \\&&\frac{A_{N,\hspace{.5mm}\gamma_{KLS}}}{(2N + 3)[(2N + 1)(1 - \gamma_{KLS}) + 2(1 - 3\gamma_{KLS})]}\epsilon ,\nonumber \\&&
\end{eqnarray}
\begin{eqnarray}
&& A_{N,\hspace{.5mm}\gamma_{KLS}} = \nonumber \\&& (2N + 3)[R_{N,\hspace{.5mm}\gamma_{KLS}}/6(1 - \gamma_{KLS}) + 2N(1 - \gamma_{KLS})] - \nonumber \\&&   [(2N + 1)(1 - \gamma_{KLS}) + 2(1 - 3\gamma_{KLS})](R_{N}/6 + 2N),\nonumber \\&&
\end{eqnarray}
\begin{eqnarray}
&& R_{N,\hspace{.5mm}\gamma_{KLS}} = -[(2N - 3)(1 - \gamma_{KLS}) + 4\gamma_{KLS}] + \nonumber \\&&  \sqrt{[(2N - 3)(1 - \gamma_{KLS}) + 4\gamma_{KLS}]^{2} + 144N/(1 - \gamma_{KLS})},\nonumber \\&& 
\end{eqnarray}
\begin{eqnarray}
R_{N} = \lim_{\gamma_{KLS}\rightarrow 0} R_{N,\hspace{.5mm}\gamma_{KLS}}.
\end{eqnarray}
In Ref. \cite{PhysRevD.96.096010}, $\eta_{\psi}$, $\eta_{\phi}$ and $\nu$ were evaluated up to four-loop level.

\subsection{$\gamma_{KLS}$-short- and long-range directed percolation}

\par Consider $\gamma_{KLS}$-short- and $\gamma_{KLS}$-long-range directed percolation \cite{JANSSEN2005147,Tauber_2005} in $d = 4 - \epsilon$ and $d = 2\sigma - \varepsilon$ dimensions, respectively. Then we obtain 
%$\eta_{\gamma} = \eta - \frac{\gamma}{1 - 3\gamma}\frac{\epsilon}{3}$, $\eta_{\sigma ,\hspace{.5mm}\gamma} = \eta_{\sigma} - \frac{\gamma}{1 - 3\gamma}\frac{2\varepsilon}{7}$, $\nu_{\gamma} = \nu + \frac{\gamma}{1 - 3\gamma}\frac{\epsilon}{8}$, $\nu_{\sigma ,\hspace{.5mm}\gamma} = \nu_{\sigma} + \frac{\gamma}{1 - 3\gamma}\frac{4\varepsilon}{7\sigma^{2}}$, $z_{\gamma} = z - \frac{\gamma}{1 - 3\gamma}\frac{\epsilon}{6}$, $z_{\sigma ,\hspace{.5mm}\gamma} = z_{\sigma} - \frac{\gamma}{1 - 3\gamma}\frac{2\varepsilon}{7}$. 
\begin{eqnarray}
\eta_{\gamma_{KLS}} = \eta - \frac{\gamma_{KLS}}{1 - 3\gamma_{KLS}}\frac{\epsilon}{3},   
\end{eqnarray}
\begin{eqnarray}
\eta_{\sigma ,\hspace{.5mm}\gamma_{KLS}} = \eta_{\sigma} - \frac{\gamma_{KLS}}{1 - 3\gamma_{KLS}}\frac{2\varepsilon}{7},  
\end{eqnarray}
\begin{eqnarray}
\nu_{\gamma_{KLS}} = \nu + \frac{\gamma_{KLS}}{1 - 3\gamma_{KLS}}\frac{\epsilon}{8}, 
\end{eqnarray}
\begin{eqnarray}
\nu_{\sigma ,\hspace{.5mm}\gamma_{KLS}} = \nu_{\sigma} + \frac{\gamma_{KLS}}{1 - 3\gamma_{KLS}}\frac{4\varepsilon}{7\sigma^{2}},
\end{eqnarray}
\begin{eqnarray}
z_{\gamma_{KLS}} = z - \frac{\gamma_{KLS}}{1 - 3\gamma_{KLS}}\frac{\epsilon}{6}, 
\end{eqnarray}
\begin{eqnarray}
z_{\sigma ,\hspace{.5mm}\gamma_{KLS}} = z_{\sigma} - \frac{\gamma_{KLS}}{1 - 3\gamma_{KLS}}\frac{2\varepsilon}{7}.
\end{eqnarray}
The standard critical exponents have been computed up to two- and one-loop level in Ref. \cite{JANSSEN2005147}, respectively.

\subsection{$\gamma_{KLS}$-short- and long-range dynamic isotropic percolation}

\par Now consider $\gamma_{KLS}$-short- and $\gamma_{KLS}$-long-range dynamic isotropic percolation \cite{JANSSEN2005147,Tauber_2005} at $d = 6 - \epsilon$ and $d = 3\sigma - \varepsilon$ dimensions, respectively. Their critical exponents are given by 
%$\eta_{\gamma} = \eta - \frac{\gamma}{1 - 3\gamma}\frac{\epsilon}{21}$, $\eta_{\sigma ,\hspace{.5mm}\gamma} = \eta_{\sigma} - \frac{\gamma}{1 - 3\gamma}\frac{3\varepsilon}{4}$, $\nu_{\gamma} = \nu + \frac{\gamma}{1 - 3\gamma}\frac{5\epsilon}{42}$, $\nu_{\sigma ,\hspace{.5mm}\gamma} = \nu_{\sigma} + \frac{\gamma}{1 - 3\gamma}\frac{\varepsilon}{2\sigma^{2}}$, $z_{\gamma} = z - \frac{\gamma}{1 - 3\gamma}\frac{\epsilon}{3}$, $z_{\sigma ,\hspace{.5mm}\gamma} = z_{\sigma} - \frac{\gamma}{1 - 3\gamma}\frac{3\varepsilon}{8}$.
\begin{eqnarray}
\eta_{\gamma_{KLS}} = \eta - \frac{\gamma_{KLS}}{1 - 3\gamma_{KLS}}\frac{2\epsilon}{21}, 
\end{eqnarray}
\begin{eqnarray}
\eta_{\sigma ,\hspace{.5mm}\gamma_{KLS}} = \eta_{\sigma} - \frac{\gamma_{KLS}}{1 - 3\gamma_{KLS}}\frac{3\varepsilon}{4}, 
\end{eqnarray}
\begin{eqnarray}
\nu_{\gamma_{KLS}} = \nu + \frac{\gamma_{KLS}}{1 - 3\gamma_{KLS}}\frac{5\epsilon}{42}, 
\end{eqnarray}
\vspace{.5cm}
\begin{eqnarray}
\nu_{\sigma ,\hspace{.5mm}\gamma_{KLS}} = \nu_{\sigma} + \frac{\gamma_{KLS}}{1 - 3\gamma_{KLS}}\frac{\varepsilon}{2\sigma^{2}},
\end{eqnarray}
\begin{eqnarray}
z_{\gamma_{KLS}} = z - \frac{\gamma_{KLS}}{1 - 3\gamma_{KLS}}\frac{\epsilon}{3}, 
\end{eqnarray}
\begin{eqnarray}
z_{\sigma ,\hspace{.5mm}\gamma_{KLS}} = z_{\sigma} - \frac{\gamma_{KLS}}{1 - 3\gamma}\frac{3\varepsilon}{8}.
\end{eqnarray}
The nongeneralized indices were computed up to two- and one-loop level in Ref. \cite{JANSSEN2005147}, respectively.

%\par \textit{Conclusions}.---We have experimentally validated NSFT by comparing its predictions with experimental measurements for manganites. The agreement was satisfactory, once the margin of error was $< 5\%$ for the most of manganites employed. NSFT is the first field-theoretic approach that proposes to describe interacting complex systems, for our knowledge, and we hope that it can shed light on future theoretical, computational and experimental studies on complex systems.

\section{Conclusions}

\par In summary, we have introduced some field-theoretic approach for evaluated the critical properties of $\gamma_{KLS}$-generalized systems undergoing continuous phase transitions, called $\gamma_{KLS}$-statistical field theory. Although it was capable of encompassing nonconventional critical exponents for real imperfect systems not described by standard statistical field theory, it is not complete. For example, it does not explain the results of Table  \ref{tableexponentsNSFT} for some other manganites, being explained for NSFT of Ref. [21] (there might be alternative models capable of accurately representing the experimental data). So, it has to be discarded for statistical mechanics generalization purposes.

\section*{Declaration of competing interest}

\par The authors declare that they have no known competing financial interests or personal relationships that could have appeared to influence the work reported in this paper.

\section*{Acknowledgments}

\par PRSC would like to thank the Brazilian funding agencies CAPES and CNPq (Grant: Produtividade 307982/2019-0) for financial support.

\bibliography{apstemplate}

\end{document}